\documentstyle[12pt]{article}

\def\be{\begin{equation}}
\def\ee{\end{equation}}
\def\MPL #1 #2 #3 {Mod.~Phys.~Lett.~{\bf#1}\ (#2) #3}
\def\NPB #1 #2 #3 {Nucl.~Phys.~{\bf#1}\ (#2) #3}
\def\PLB #1 #2 #3 {Phys.~Lett.~{\bf#1}\ (#2) #3}
\def\PR #1 #2 #3 {Phys.~Rep.~{\bf#1}\ (#2) #3}
\def\PRD #1 #2 #3 {Phys.~Rev.~{\bf#1}\ (#2) #3}
\def\PRL #1 #2 #3 {Phys.~Rev.~Lett.~{\bf#1}\ (#2) #3}
\def\RMP #1 #2 #3 {Rev.~Mod.~Phys.~{\bf#1}\ (#2) #3}
\def\ZP #1 #2 #3 {Z.~Phys.~{\bf#1}\ (#2) #3}

\begin{document}

\begin{titlepage}

\title{{\bf On the spontaneous CP breaking in the Higgs sector of the
Minimal Supersymmetric Standard Model}
\thanks{Work partly supported by CICYT under
contract AEN93--0139.}}

\author{
{\bf J.R. Espinosa} \thanks{Supported by a Comunidad de Madrid Grant.}\ ,
{\bf J.M. Moreno} and  {\bf M. Quir\'os} \\
Instituto de Estructura de la Materia, CSIC\\
Serrano 123, E--28006 Madrid, Spain}

\date{}
\maketitle
\vspace{1.5cm}
\def\baselinestretch{1.15}
\begin{abstract}

We revise a recently proposed mechanism for spontaneous CP breaking at
finite temperature in the Higgs sector of the Minimal Supersymmetric
Standard Model, based on the contribution of squarks, charginos and
neutralinos to the one-loop effective potential. We have included plasma
effects for all bosons and added the contribution of neutral scalar and
charged Higgses. While the former have little effect, the latter provides
very strong extra constraints on the parameter space and change drastically
the previous results. We find that CP can be spontaneously broken at the
critical temperature of the electroweak phase transition without any
fine-tuning in the parameter space.

\end{abstract}

\thispagestyle{empty}

\vskip-17.5cm
\rightline{{\bf hep-ph/9308315}}
\rightline{{\bf IEM--FT--76/93}}
\rightline{{\bf July 1993}}
\vskip3in

\end{titlepage}

\def\baselinestretch{1.1}

{\bf 1.} The issue of spontaneous CP breaking (SCPB) at zero temperature
has been recently proposed in supersymmetric theories as an appealing
alternative \cite{Barr} to the Peccei-Quinn mechanism to solve
the strong CP problem, and also as a natural explanation for the
smallness of CP violating phases \cite{Pomarol1} contributing to the
neutron electron dipole moment (NEDM). In the minimal
supersymmetric standard model (MSSM), SCPB at zero temperature
can be triggered by radiative corrections \cite{Maekawa}.
Unfortunately, as established on general grounds by Georgi and Pais
\cite{Georgi}, it requires the existence of a Higgs boson with a mass
of a few GeV \cite{Pomarol2}, which has been ruled out at LEP \cite{LEP}.

On the other hand, CP violation at finite temperature is one of the
key ingredients to generate baryon asymmetry at the electroweak phase
transition \cite{Kuzmin}. It has been recently realized \cite{Comelli1}
that temperature effects can trigger SCPB in the MSSM at the critical
temperature of the electroweak phase transition, and baryon asymmetry
with the help of tiny CP violating phases giving rise to a NEDM well below
its present experimental limit \cite{Comelli2}. The analysis of
Refs. \cite{Comelli1,Comelli2} was based on the contribution of squarks,
charginos and neutralinos to the one-loop effective potential at finite
temperature.

In this paper we have analyzed SCPB in the MSSM at finite temperature,
taking into account plasma effects for all bosons and including the
contribution of the Higgs sector to the effective potential. We have found
that plasma effects have little influence on the SCPB, but the contribution
from neutral scalar and charged Higgses changes drastically the results
found in Ref. \cite{Comelli1,Comelli2} giving rise to a much stronger
constraint on the parameter space. Nevertheless we find that SCPB can
be accomplished without any fine-tuning of the parameters.

\vspace{1cm}
{\bf 2.} The most general scalar potential for the
two--Higgs doublet model that is renormalizable
and $SU(2)\times U(1)$ invariant is given by \cite{Higgshunter}
\be
\begin{array}{c}
V = {m_{1}}^{2}|H_1|^2 + {m_{2}}^{2}|H_2|^2 - ({m_{3}}^{2} H_1 H_2 +
h.c.) +\lambda_1 |H_1|^4 +\lambda_2 |H_2|^4 + \lambda_3 |H_1|^2 |H_2|^2
\vspace{.3cm}\\
+\lambda_4|H_1 H_2|^2
+(\lambda_5 (H_1 H_2)^2  + \lambda_6 |H_1|^2 H_1 H_2 +\lambda_7 |H_2|^2
H_1 H_2 + h.c.),
\end{array}
\ee
where ${m_3}^2$, $\lambda_5$, $\lambda_6$ and $\lambda_7$ are taken real,
so that there is no explicit CP violation at tree level.
However, SCPB can be triggered if
the vacuum expectation
values of the neutral components of the Higgs fields get a relative
phase $\delta\neq n\pi$, with integer $n$.
(We can take $\langle H_1^o\rangle=v_1$,
$\langle H_2^o\rangle=v_2 \: e^{i\delta}$ without any loss of generality.)
This requires,
\be
\begin{array}{c}
\lambda_5>0,\vspace{.3cm}\\
-1<\cos \delta = \displaystyle{\frac{m_3^2-
\lambda_6 v_1^2 -\lambda_7 v_2^2}
{4 \lambda_5 v_1 v_2}}< 1.
\end{array}
\ee

In the MSSM these conditions are not satisfied
(at tree level) because in this case supersymmetry gives
\be
\begin{array}{cl}
\lambda_1 &= \lambda_2\:\:=\:\:\frac{1}{8}({g}^{2}+{g'}^{2}),\vspace{.3cm}\\
\lambda_3 &= \frac{1}{4}({g}^{2}-{g'}^{2}),\vspace{.3cm}\\
\lambda_4 &= -\frac{1}{2}{g}^2, \vspace{.3cm}\\
\lambda_5 &= \lambda_6\:\:=\:\:\lambda_7\:\:=\:\:0,
\end{array}
\ee
where $g$ and $g'$ are the $SU(2)_L$ and $U(1)_Y$ gauge couplings.
Of course, after supersymmetry breaking, one--loop effects
change these couplings and there are \cite{Maekawa} finite
contributions (proportional to the soft parameters) that can produce
SCPB.
At finite temperature, there will be additional corrections coming
from the thermal modes running in the loops.

We can write the one--loop effective potential, including the zero
temperature terms, as
\be
\Delta V_1 = T^4 \left( \sum_b g_b V_b -  \sum_f g_f V_f\right),
\ee
where $g_b$ ($g_f$) denotes the number of degrees of freedom of
bosons (fermions). In the bosonic part we sum over stops, neutral scalar
and charged Higgses, and Goldstone bosons ($b=\tilde{t}_{L,R},
h,H,H^{\pm},G^{\pm}$), and in the fermionic one we consider
neutralinos and charginos ($f=\tilde{\chi}^o,\tilde{\chi}^{\pm}$).
Working in the 't Hooft--Landau gauge and in the
$\overline{{\rm DR}}$ renormalization scheme, we will decompose the
bosonic components as
\be
V_b=V_{nz}(y_b^2)+V_{z}(\bar{y}_b^2).
\ee
$V_{nz}$ includes the bosonic zero temperature part plus
the finite temperature contribution coming from $n\neq0$ Matsubara
modes \cite{DJW},
\be
V_{nz}(y_b^2)=Tr\left\{\displaystyle{\frac{y_b^4}{64\pi^2}}
\left[\log\left(\frac{T^2y_b^2}{Q^2}\right) -
\displaystyle{\frac{3}{2}}\right] + \displaystyle{\frac{1}{2\pi^2}}J_b(y_b^2)
\right\},
\ee
where $y_b^2={\cal M}_b^2/T^2$ is the bosonic mass matrix rescaled with
temperature, $Q$ is the renormalization scale, and the finite
temperature one--loop term is given by the usual integral expresion
with the $n=0$ part subtracted out,
\be
J_b(y_b^2)=\int_{0}^{\infty}dx\ x^2\
\log\left(1-e^{-\sqrt{x^2+y_b^2}} \right) -
\left(-\displaystyle\frac{\pi}{6}y_b^3 \right).
\ee
The $n=0$ term is given by \cite{DJW}
\be
V_{z}(\bar{y}_b^2)
=Tr\left\{-\displaystyle{\frac{\bar{y}_b^3}{12\pi}} \right\},
\ee
and now $\bar{y}_b^2=({\cal M}_b^2+\Pi_b)/T^2$, where $\Pi_b$ is the thermal
polarization matrix for bosons,
contains the Debye masses \cite{Gross}.

As the fermions do not have zero Matsubara modes, $V_f$ is just
given by \cite{DJW}
\be
V_f=Tr\left\{\displaystyle{\frac{y_f^4}{64\pi^2}}
\left[\log\left(\frac{T^2y_f^2}{Q^2}\right) -
\displaystyle{\frac{3}{2}}\right] + \displaystyle{\frac{1}{2\pi^2}}J_f(y_f^2)
\right\},
\ee
with
\be
J_f(y_f^2)=\int_{0}^{\infty}dx\ x^2\
\log\left(1+e^{-\sqrt{x^2+y_f^2}} \right) ,
\ee
and $y_f^2={\cal M}_f^2/T^2$, where ${\cal M}_f^2$ is the
fermionic mass matrix.

The bosonic and fermionic mass matrices
are functions of the classical fields
$H_1^o$, $H_2^o$, $\bar{H_1}^o$ and $\bar{H_2}^o$, and one can expand the
potential (4) in powers of them to obtain the one--loop
corrections to the tree level couplings of Eq. (1).
Assuming as in Ref. \cite{Comelli1} equal soft supersymmetry
breaking masses for left and right-handed squarks
\footnote{This is just a {\it technical} simplification
such that the mass matrix at the origin, $H_1^o=H_2^o=0$, is
proportional to the identity matrix.}
$m_Q^2=m_U^2\equiv
\tilde{m}^2$ one can
easily find the contributions from stop loops to $m_3^2$, $\lambda_5$,
$\lambda_6$ and $\lambda_7$:
\begin{eqnarray}
\Delta^{(s)} {m_3^2} &=&
-6 h_t^2 A_t \mu I'_{\tilde{t}},
\vspace{.3cm}\\
\Delta^{(s)} \lambda_5&=&
{\displaystyle\frac{1}{2}}h_t^4
{\displaystyle\frac{A_t^2\mu^2}{T^4}} I'''_{\tilde{t}}, \vspace{.3cm}\\
\Delta^{(s)} \lambda_6&=&
h_t^2{\displaystyle\frac{A_t\mu}{T^2}}\left[
{\displaystyle\frac{3}{4}}(g^2+g'^2)
I''_{\tilde{t}}
+h_t^2{\displaystyle\frac{\mu^2}{T^2}}I'''_{\tilde{t}}\right],
\vspace{.3cm}\\
\Delta^{(s)} \lambda_7&=&
h_t^2{\displaystyle\frac{A_t\mu}{T^2}}\left[
\left(6 h_t^2 -{\displaystyle\frac{3}{4}}(g^2+g'^2) \right)
I''_{\tilde{t}}
+h_t^2{\displaystyle\frac{A_t^2}{T^2}}I'''_{\tilde{t}}
\right],
\end{eqnarray}
where \footnote{The high-$T$ expansion of Eqs. (11-14) coincides with
Eqs. (15-18) in Ref. \cite{Comelli1}, except that we find 1/512 instead
of 1/256 in Eq. (16) of \cite{Comelli1}.}
$\mu$ is the supersymmetric Higgs mixing mass, $A_t$ the trilinear
soft breaking parameter corresponding to $h_t Q \cdot H_2 U$ in the
superpotential,
\begin{eqnarray}
I_b&\equiv & I_z+I_{nz}\equiv {\displaystyle\frac{d}{d\bar{y}_b^2}}
V_{z}(\bar{y}_b^2)+{\displaystyle\frac{d}{dy_b^2}}
V_{nz}(y_b^2), \vspace{.3cm}\\
I'_b&\equiv &I'_z+I'_{nz}\equiv {\displaystyle\frac{d}{d\bar{y}_b^2}}
I_{z}(\bar{y}_b^2)+ {\displaystyle\frac{d}{dy_b^2}}
I_{nz}(y_b^2)
\end{eqnarray}
(and a similar definition for higher derivatives),
and now $y_{\tilde{t}}^2=\tilde{m}^2/T^2,\:
\bar{y}_{\tilde{t}}^2=(\tilde{m}^2+\Pi_{\tilde{t}})/T^2$, with
\cite{eqz3,beqz}
\be
\Pi_{\tilde{t}}\simeq {\displaystyle\frac{2}{3}}g_s^2 T^2.
\ee
where $g_s$ is the $SU(3)$ gauge coupling.
We are neglecting in Eq. (17) contributions of order
$g^2,g'^2$ and $h_t^2$ compared to the $g_s^2$ part.
In this way the field independent mass
matrix for stops with the thermal screening included is
proportional to the identity matrix and the
expressions (11-14) apply.

The mass matrices of neutral and
charged Higgses at the origin, $H_1^o=H_2^o=0$,
are never proportional to the identity
and so the corresponding one--loop corrections
involve finite differences instead of derivatives of the $I_b$
functions. In particular,
\begin{eqnarray}
\Delta^{(h)} {m_3^2} &=&
-{\displaystyle\frac{1}{2}}\left\{(g^2+g'^2)+ 2g^2\right\} m_3^2\Delta I_h,
\vspace{.3cm}\\
\Delta^{(h)} \lambda_5&=&
{\displaystyle\frac{1}{48}}{\displaystyle\frac{m_3^4}{T^4}}
\left\{(g^2+g'^2)^2+2g^4\right\}  \Delta_3 I_h,  \vspace{.3cm}\\
\Delta^{(h)} \lambda_6&=&
\left.{\displaystyle\frac{1}{8}}{\displaystyle\frac{m_3^2}{T^2}}
 \right[\left\{(g^2+g'^2)^2+2 g^4\right\}\Delta I'_h
\vspace{.2cm}\\
& & \left.+\left\{(g^2+g'^2)^2+ g^2 g'^2 \right\}
{\displaystyle\frac{m_1^2-m_2^2}{3 T^2}}\Delta_3 I_h\right],
\vspace{.3cm} \nonumber \\
\Delta^{(h)} \lambda_7&=&
\left.{\displaystyle\frac{1}{8}}{\displaystyle\frac{m_3^2}{T^2}}
 \right[\left\{(g^2+g'^2)^2+2 g^4\right\}\Delta I'_h
\vspace{.2cm}\\
& & \left.-\left((g^2+g'^2)^2+ g^2 g'^2\right)
{\displaystyle\frac{m_1^2-m_2^2}{3 T^2}}\Delta_3 I_h\right], \nonumber
\end{eqnarray}
where the contribution to $\Delta^{(h)}$ proportional to
$(g^2+g'^2)$ inside the curly brackets
comes from the neutral scalar Higgs sector, and the
rest, proportional to $g^2$, from the charged Higgs sector.
$\Delta I_h$ and $\Delta_3 I_h$ in Eqs. (18-21) are defined by
\be
\begin{array}{cl}
\Delta I_h&\equiv {\displaystyle
\frac{I_{nz}(y_+^2)-I_{nz}(y_-^2)}{y_+^2-y_-^2}}
+{\displaystyle
\frac{I_{z}(\bar{y}_+^2)-I_{z}(\bar{y}_-^2)}{\bar{y}_+^2-\bar{y}_-^2}},
\vspace{.3cm}\\
\Delta_3 I_h&\equiv {\displaystyle\frac{6}{(y_+^2-y_-^2)^2}}\left[
I'_{nz}(y_+^2)+I'_{nz}(y_-^2)-2{\displaystyle \frac{I_{nz}(y_+^2)
-I_{nz}(y_-^2)}{y_+^2-y_-^2} } \right]  \vspace{.5cm}\\
& +{\displaystyle\frac{6}{(\bar{y}_+^2-\bar{y}_-^2)^2}}\left[
I'_{z}(\bar{y}_+^2)+I'_{z}(\bar{y}_-^2)-
2{\displaystyle \frac{I_{z}(\bar{y}_+^2)
-I_{z}(\bar{y}_-^2)}{\bar{y}_+^2-\bar{y}_-^2} } \right],
\end{array}
\ee
and $y_{\pm}^2=m_{\pm}^2/T^2,\:\: \bar{y}_{\pm}^2=\bar{m}_{\pm}^2/T^2$ as
usual, with $m_{\pm}^2$ the
Higgs masses for zero classical fields:
\be
m_{\pm}^2={\displaystyle\frac{1}{2}}\left\{ m_1^2+m_2^2 \pm
\sqrt{(m_2^2-m_1^2)^2 +4 m_3^2} \right\},
\ee
and $\bar{m}_{\pm}^2$ the corresponding masses
in the presence of screening:
\be
\bar{m}_{\pm}^2={\displaystyle\frac{1}{2}}\left\{ \bar{m}_1^2+\bar{m}_2^2 \pm
\sqrt{(\bar{m}_2^2-\bar{m}_1^2)^2 +4 m_3^2} \right\},
\ee
where $m_1^2$ and $m_2^2$ are functions of
$m_3^2$, $\tan \beta$ and the other parameters of the theory,
by means of the zero-temperature (including the one-loop corrections)
minimization conditions, and \cite{eqz3,beqz}
\begin{eqnarray}
\bar{m}_1^2&=&m_1^2 +
\Pi_{h_1}=m_1^2+{\displaystyle\frac{1}{8}}(3 g^2 + g'^2)T^2,
\vspace{.3cm}\\
\bar{m}_2^2&=&m_2^2 +
\Pi_{h_2}=m_2^2+{\displaystyle\frac{1}{8}}(3 g^2 + g'^2 +6 h_t^2)T^2.
\end{eqnarray}

We can also assume that $^1$ $\mu^2=M_1^2=M_2^2$
(where $M_2$ and $M_1$ are the soft Majorana masses for
$SU(2)_L$ and $U(1)_Y$ gauginos, respectively)
and then find, for charginos:
\begin{eqnarray}
\Delta^{(c)} {m_3^2} &=&
-{\rm sign}(\mu)4g^2\mu^2 I'_{\chi},
\vspace{.3cm}\\
\Delta^{(c)} \lambda_5&=&
{\displaystyle-\frac{1}{3}}g^4
{\displaystyle\frac{\mu^4}{T^4}} I'''_{\chi}, \vspace{.3cm}\\
\Delta^{(c)} \lambda_6&=&\Delta^{(c)} \lambda_7=
{\rm sign}(\mu)g^4{\displaystyle\frac{2}{3}}
{\displaystyle\frac{\mu^2}{T^2}}\left[
3I''_{\chi} +{\displaystyle\frac{\mu^2}{T^2}}I'''_{\chi}\right],
\end{eqnarray}
and for neutralinos
\footnote{We correct here Eq. (26) of Ref. \cite{Comelli1}
where the proportionality coefficient of $\Delta^{(c)}
\lambda_{5,6,7}$ was written as $(g^2+g'^2)/2g^2$.}:
\begin{eqnarray}
\Delta^{(n)} {m_3^2} &=&{\displaystyle\frac{g^2+g'^2}{2g^2}}
\Delta^{(c)} {m_3^2},\vspace{.3cm}\\
\Delta^{(n)} \lambda_{5,6,7}&=&
{\displaystyle\frac{(g^2+g'^2)^2}{2g^4}}\Delta^{(c)}\lambda_{5,6,7},
\end{eqnarray}
where now $y_{\chi}^2=\mu^2/T^2$ and all the integrals are fermionic.
Derivatives are defined, as usual,
\be
I_f=V'_f\equiv  {\displaystyle\frac{d}{dy_f^2}}
V_f(y_f^2).
\ee

\vspace{1cm}
{\bf 3.} We analyze now the effect described in the
previous section. The parameter $m_3^2$
will be traded by $m_A^2$, the
physical mass of the pseudoscalar which includes all radiative corrections
(at zero--temperature). None of conditions (2) is satisfied at tree level
(in particular, $\lambda_5=0$ and $\delta =n \pi$),
neither after including one--loop radiative corrections at zero--temperature,
if we fix $m_A$ beyond its present experimental limit \cite{Pomarol2}.
However, they can be satisfied when radiative corrections at
finite--temperature are included.
In particular, $\lambda_a,\ (a=5,6,7)$ acquires a renormalized value
at finite temperature
\be
\lambda_a(T)=\sum_{i=s,h,c,n}\Delta^{(i)}\lambda_a
\ee
as given by Eqs. (12-14), (19-21), (28-29) and (31). Also $m_3^2$ gets
renormalized as
\be
m_3^2(T)=m_3^2+\sum_{i=s,h,c,n}\Delta^{(i)}m_3^2
\ee
from Eqs. (11), (18), (27) and (30). We have computed, in (33) and (34),
$\Delta^{(s,c,n)}$ numerically, from the integrals (7) and (10), while
$\Delta^{(h)}$ is computed using its finite temperature expansion, because
in the Higgs sector the corresponding value of the integral (7) is well
defined only through its high-$T$ expansion.

A glance at the previous section
(Eqs. (12), (19), (28) and (31)) shows that
$\Delta^{(s)}\lambda_5\le 0$, while
$\Delta^{(c,n)}\lambda_5\ge 0$, so that the
contribution of fermions (charginos and neutralinos) have to compensate
that of squarks, if no other contributions were considered,
so that one would obtain in the plane ($\mu,\tilde{m}$)
a lower bound contour, as in Ref. \cite{Comelli1}.
We have analyzed the effects of the Debye screening
on the latter contour and found it gets shifted towards smaller
($\sim 5 \%$) values of $\tilde{m}$. We can conclude from this that plasma
corrections in the squark sector are not very important.
Now the contribution from the Higgs sector
(where plasma effects are essential to make the effective potential
at $H_1^o=H_2^o=0$ real) is negative, and has an
infrared singularity ($\propto-(T/\bar{m}_-)$), responsible for the
failure of perturbative expansion \cite{DJW},
for values of ($\mu,\tilde{m}$) such that $\bar{m}_-=0$
(long-dashed lines in Figs. 1-3).
This creates in the plane ($\mu,\tilde{m}$) an upper
bound contour which severely constrains the
region of the parameter space where $\lambda_5>0$.
This behaviour is exemplified in Figs.1-3 where
$\lambda_5=0$ is plotted in the plane ($\mu,\tilde{m}$) for different
values of $T$, $m_A$ and the supersymmetric parameters
\footnote{We have checked that the theory remains perturbative,
{\it i.e.} that $\bar{\beta}_{\pm}\equiv g^2 T/\bar{m}_{\pm}<1$, in
the region where $\lambda_5>0$ in Figs. 1-3.}. We can see from
Figs. 1b, 2b and 3b that the lower and upper contours provide a combined
region where $\lambda_5>0$. In the absence of mixing $A_t=0$,
Figs. 1a, 2a and 3a, there is no
contribution coming from the squark sector and, consequently, the lower
contour disappears. Only the upper contour
(from the Higgs sector) constrains the region where
$\lambda_5>0$. If we compare the region where $\lambda_5(T)>0$ for different
temperatures, {\it e.g.} Figs. 1 and 2, we see it decreases with decreasing
T, and for any given point in the plane ($\mu,\tilde{m}$) the condition
$\lambda_5>0$ will stop being satisfied at a given temperature.

In the region where $\lambda_5(T)>0$ the
condition $\mid \cos \delta(T) \mid <1$ where
\be
\cos \delta(T) = \displaystyle{\frac{m_3^2(T)-
\lambda_6(T) v_1^2(T) -\lambda_7(T) v_2^2(T)}
{4 \lambda_5(T) v_1(T) v_2(T)}}
\ee
is also required to have SCPB.
This condition is satisfied for values of $v(T)$ ($v^2(T)=v_1^2(T)
+v_2^2(T)$) such that $v_-(T)\le v(T)\le v_+(T)$, where
\be
v_{\pm}(T)=\frac{m_3^2(T)}{\cos^2 \beta(T)}\left\{
\lambda_6(T)+\lambda_7(T) \tan^2 \beta(T) \mp 4\lambda_5(T)\tan \beta(T)
\right\}^{-1}
\ee
For a given point in the plane ($\mu,\tilde{m}$) the values of
$v_{\pm}(T)$ are very close to each other and depend on $\tan \beta(T)$.
For a first--order phase transition the values of $\tan \beta(T)$ and
$\langle v(T)\rangle$ at the critical temperature
are dynamically fixed by all the
parameters of the supersymmetric theory. In the absence of a complete
analysis of the phase transition in the MSSM \cite{beqz}, we will take
$\tan \beta(T)=\tan \beta$, which we know is a good approximation at least
for large values of $m_A$, to get an estimate of the values of
$v_{\pm}$ for the different points in the plane ($\mu,\tilde{m}$).
At very high temperatures, $\langle v(T)\rangle =0$ and the condition
(2) will not be satisfied in general for a fixed point in
the plane ($\mu,\tilde{m}$). At the critical temperature
$T_c$ the field
will go from the origin to $\langle v(T_c)\rangle$
and produce a maximal CP violation
($\Delta \delta =\pm \pi$)
if $0<v_{-}(T_c)<v_{+}(T_c)<\langle v(T_c)\rangle$ for the chosen point
in the plane ($\mu,\tilde{m}$).

We have plotted in Figs. 1-3 level contours corresponding to  different
values of $0\le v_-(T)\le T$. If the phase transition is strong enough,
{\it i.e.} $\langle v(T)\rangle \sim T$, then the whole region with
$v_-(T)\le T$ and $\lambda_5>0$ satisfies the condition of CP
violation. If the phase transition is stronger (weaker), {\it i.e.}
$\langle v(T)\rangle >T$ ($\langle v(T)\rangle <T$),
then the allowed region showed in Figs. 1-3 will
be further enlarged (reduced). For values of $\langle \tan \beta(T)
\rangle$ different from $\tan\beta$, the allowed regions in Figs. 1-3
would get distorted, but still a wide region would appear.

\vspace{1cm}
{\bf 4.} In conclusion we have estimated the region of maximal CP violation
at the critical temperature of the electroweak phase transition in the
space of supersymmetric parameters.
We have studied
for simplicity the degenerate case of $m_Q=m_U$ in the stop sector, and
$M_1=M_2=\mu$ in the gaugino/higgsino sector.
We have found that the Higgs sector
changes dramatically previous results on the same effect.
Nevertheless the mechanism can work in a wide region of the parameter
space without any fine-tuning, and trigger easily
maximal CP violation during the first order phase transition.
However this requires, as a general tendency, small values of $m_A$ and large
values of $\tan \beta$, while preliminary results on the phase transition
in the MSSM \cite{eqz3,beqz} seem to point towards the opposite direction
in order not to wash out the generated baryon asymmetry.
The analysis of the non-degenerate case
$m_Q\neq m_U$, $M_1\neq M_2\neq \mu$ (now in progress) seems necessary
to find out whether this mechanism can be realistically
used to generate baryon asymmetry
at the phase transition in the MSSM.

\section*{Acknowledgements}
We acknowledge discussions with D. Comelli and A. Riotto.

\newpage

\begin{description}
\item[Fig.1]

1a) Contours of $\bar{m}_-=0$ (long-dashed line),
$\lambda_5=0$ (horizontal solid line),
$v_-(T)/T=1$ (outer vertical solid lines) and
$v_-(T)/T=0$ (inner vertical solid lines)
for $m_t=160\ GeV$, $T=150\ GeV$,
$m_A=50\ GeV$, $\tan \beta =8$ and $A_t=0$; 1b) The same as in 1a) but for
$A_t=50\ GeV$. Here the contour of $\lambda_5=0$ is closed on the left and
the contours of $v_-(T)/T=1$ (outer vertical solid lines)
and $v_-(T)/T=0.6$ (short-dashed line) are shown.

\item[Fig.2]
The same as in Fig. 1 but with $T=100\ GeV$ and $\tan \beta=18$. In 2b)
the short-dashed line corresponds to $v_-(T)/T=0.7$

\item[Fig.3]
The same as in Fig. 1 but for $m_A=75\ GeV$ and $\tan \beta =18$. In 3b)
the short-dashed line corresponds to $v_-(T)/T=0.5$

\end{description}

\end{document}